\documentclass[twocolumn,aps,prd,amsmath,amssymb,nofootinbib]{revtex4-1}

\usepackage{graphicx,bm,tikz,amsmath}
\newcommand{\beq}{\begin{equation}}
\newcommand{\bea}{\begin{eqnarray}}
\newcommand{\eeq}{\end{equation}}
\newcommand{\eea}{\end{eqnarray}}
\newcommand{\bal}{\begin{align}}
\newcommand{\eal}{\end{align}}

\newcommand{\lp}{\left(}
\newcommand{\rp}{\right)}
\newcommand{\mpl}{m_{Pl}}

\begin{document}

\title{Detecting Dark Matter with Imploding Pulsars in the Galactic Center}

\author{Joseph Bramante}
\affiliation{Department of Physics, 225 Nieuwland Science Hall,\\ University of Notre Dame, Notre Dame, IN 46556, USA}

\author{Tim Linden}
\affiliation{Kavli Institute for Cosmological Physics 5640 South Ellis Avenue \\University of Chicago Chicago, IL 60637}

\begin{abstract}
The paucity of old millisecond pulsars observed at the galactic center of the Milky Way could be the result of dark matter accumulating in and destroying neutron stars. In regions of high dark matter density, dark matter clumped in a pulsar can exceed the Schwarzschild limit and collapse into a natal black hole which destroys the pulsar. We examine what dark matter models are consistent with this hypothesis and find regions of parameter space where dark matter accumulation can significantly degrade the neutron star population within the galactic center while remaining consistent with observations of old millisecond pulsars in globular clusters and near the solar position. We identify what dark matter couplings and masses might cause a young pulsar at the galactic center to unexpectedly extinguish. Finally, we find that pulsar collapse age scales inversely with the dark matter density and linearly with the dark matter velocity dispersion. This implies that maximum pulsar age is spatially dependent on position within the dark matter halo of the Milky Way. In turn, this pulsar age spatial dependence will be dark matter model dependent.
\end{abstract}

\maketitle

Dark matter (DM) is evident in the rotational velocities of galaxies, the equation of state of the primordial universe, and the gravitational lensing of colliding galactic clusters. Although its gravitational characteristics are well established, its other putative, impuissant interactions have only been constrained. Nevertheless, there are hints of DM couplings in gamma ray excesses from the galactic center (GC)~\citep{Hooper:2010mq, Abazajian:2012pn, Gordon:2013vta, Daylan:2014rsa} a keV photon line in galactic clusters~\citep{Bulbul:2014sua,Boyarsky:2014jta}, and the cored out mass profiles of dwarf galaxies~\citep{Rocha:2012jg, Peter:2012jh, Vogelsberger:2012ku, Zavala:2012us}. 

Recently, X-ray observations by the NuSTAR and Swift satellites observed pulsating emission from a magnetar located at an angular distance of only 3" from the dynamical center of the galaxy -- corresponding to a three-dimensional separation of only $\sim0.1~{\rm pc}$~\citep{Mori:2013yda, Kennea:2013dfa}. This finding was followed-up by several groups, who found radio pulsations from the same object~\cite{2013ATel.5043....1E, 2013ATel.5053....1B}. The radio data indicated that the temporal broadening of the pulsar beam by electrons in the GC region was rather small, compared to theoretical predictions~\citep{Spitler:2013uva}. Previously, it had been reasonable to assume that the temporal broadening at radio energies was larger than the pulse frequency, making it impossible to observe even bright pulsars located in the GC. These new measurements of the temporal broadening instead indicate that a significant fraction of pulsars, and even millisecond pulsars (MSPs), located in the GC should be detectable by current radio surveys~\citep{Bower:2013tva}. The null detection of additional pulsars thus creates significant tension between models of binary pulsar evolution and current observations, and is now termed the ``missing pulsar problem"~\cite{Pfahl:2003tf, Macquart:2010vf, Dexter:2013xga}. While this tension is extremely statistically significant, \citep{Chennamangalam:2013zja} show that it is potentially resolvable through alterations in the luminosity function of the pulsar population. 

In this work we propose that DM may scatter into dense cores in pulsars and exceed the Schwarzschild limit within the lifetime of GC pulsars, especially MSPs. Indeed, many bounds have been placed on non-annihilating (a.k.a. Asymmetric) DM with the observation of old and compact astrophysical objects \cite{Goldman:1989nd,Kouvaris:2007ay,McCullough:2010ai,deLavallaz:2010wp,Kouvaris:2010vv,Kouvaris:2010jy,McDermott:2011jp,Kouvaris:2011fi,Kouvaris:2011gb,Casanellas:2012jp,Bramante:2013hn,Bell:2013xk,Goldman:2013qla,Jamison:2013yya,Bertoni:2013bsa,Bramante:2013nma,Kouvaris:2013kra}. We note that the DM NS collapse mechanism requires that the DM annihilation rate be nearly non-existent, and that the symmetry enforcing this remains valid during collapse to a black hole. Otherwise the DM will annihilate away before crossing the Schwarzschild radius \cite{Bramante:2013hn,Bell:2013xk}.

The bounds derived from pulsar collapse require a sequence of calculations in order to determine whether surrounding DM collapses pulsars: (i) DM accumulates in the pulsar via gravitational infall followed by DM-fermion scatter, (ii) the DM thermalizes into a core at the center of the pulsar, (iii) for bosonic DM a Bose-Einstein condensate forms, while for fermionic DM a substantial attractive self-interaction permits collapse, (iv) the DM energy must be minimized at arbitrarily small distances so DM collapse proceeds, (v) the resulting black hole (BH) must accrete circumferential material quickly or it will evaporate via Hawking radiation. Below we quantify these calculations and utilize them to determine what mass and couplings DM would have if a radially dependent maximum pulsar age at the GC is a consequence of DM pulsar destruction. Particularly, we demonstrate that a single precipitous extinction of a young pulsar at the GC could fix a mass/fermion coupling ratio for DM.

In the treatment that follows, we assume asymmetric DM with a small scattering interaction with nucleons via a heavy, decoupled mediator. For bosonic DM, these are the only necessary requirements, though we also consider the affect of a small ($\lambda=10^{-15}$) quartic coupling. For fermionic DM, it is necessary to include a large attractive self-interaction to precipitate collapse in spite of Fermi degeneracy pressure -- here this is implemented with a Yukawa coupling to a light boson. For a fuller treatment of asymmetric dark matter, see \cite{Petraki:2013wwa,Zurek:2013wia} and references therein.

If DM has a scattering interaction with nucleons (or electrons \cite{Bertoni:2013bsa}), then ambient DM will collect in pulsars. A detailed calculation for the scattering of circumferential DM on pulsars \cite{Gould:1987ir,Kouvaris:2007ay,Bell:2013xk,Bramante:2013nma} determines that the total capture rate is given by
\begin{align}
C_{X} \simeq \sqrt{\frac{6}{\pi}}  \lp\frac{\rho_X}{\bar{v}_X} \rp \frac{\xi N_B v_{\rm esc}^2}{m_X}  \left[1- \frac{1-{\rm exp}(-B^2)}{B^2}\right] f \lp \sigma_{nX} \rp
\label{eq:scattering}
\end{align}
Here the local DM density is $\rho_X$, the velocity dispersion is $\bar{v}_X$, the number of nucleons in the pulsar is $N_B$, the pulsar escape velocity is $v_{\rm esc}$, the DM mass is $m_X$, and $f \lp \sigma_{nX} \rp=\sigma_{\rm sat}\lp 1-{\rm exp} (-\sigma_{nX}/ \sigma_{\rm sat}) \rp$ is a function of the DM-nucleon scattering cross section $\sigma_{nX}$. Equation \ref{eq:scattering} includes refinements that take into account a maximum DM-pulsar cross section $\sigma_{\rm sat} \equiv R_{NS}^2/(0.45 N_B \xi)$ where $R_{NS}$ is the pulsar radius. The maximum scattering cross-section of the pulsar along with the capture rate depend on Pauli blocking, that is whether incoming DM can excite nucleons above the Fermi surface $\xi \sim {\rm min} [\frac{m_X}{{0.2 \rm GeV}},1]$ (see \cite{McDermott:2011jp,Bell:2013xk,Bertoni:2013bsa,Bramante:2013nma}). Finally, the term in square brackets in  (\ref{eq:scattering}) accounts for DM too energetic to be captured by nucleon scattering,
$
B^{2} = \frac{6 \, v_{\rm esc}^{2}}{\bar{v}_X^2}\frac{m_{X} m_{B}}{(m_{X}-m_{B})^{2}}.
$
For a nucleon mass $m_{B} \sim {\rm GeV}$ this will be negligible until the DM is quite heavy, $m_X \gtrsim 10^6~ {\rm GeV}$.

The parenthetical term of (\ref{eq:scattering}) is the most salient feature of the capture rate in this study. Capture scales as DM density over velocity, and thus a 7 Gyr old pulsar near the solar position, where $\frac{\rho_X}{\bar{v}_X}=\frac{0.4 ~{\rm GeV/cm^3}}{220 ~{\rm km/s}}$ has a smaller captured DM mass than a 40 kyr pulsar located 0.1~pc from the GC, where $\frac{\rho_X}{\bar{v}_X}\sim\frac{7 \times 10^4 ~{\rm GeV/cm^3}}{200 ~{\rm km/s}}$. Indeed, these separate observations will provide close upper and lower bounds resulting in a well-defined relation between DM mass and nucleon scattering.

\begin{figure}
\begin{tabular}{c}
\includegraphics[scale=.6]{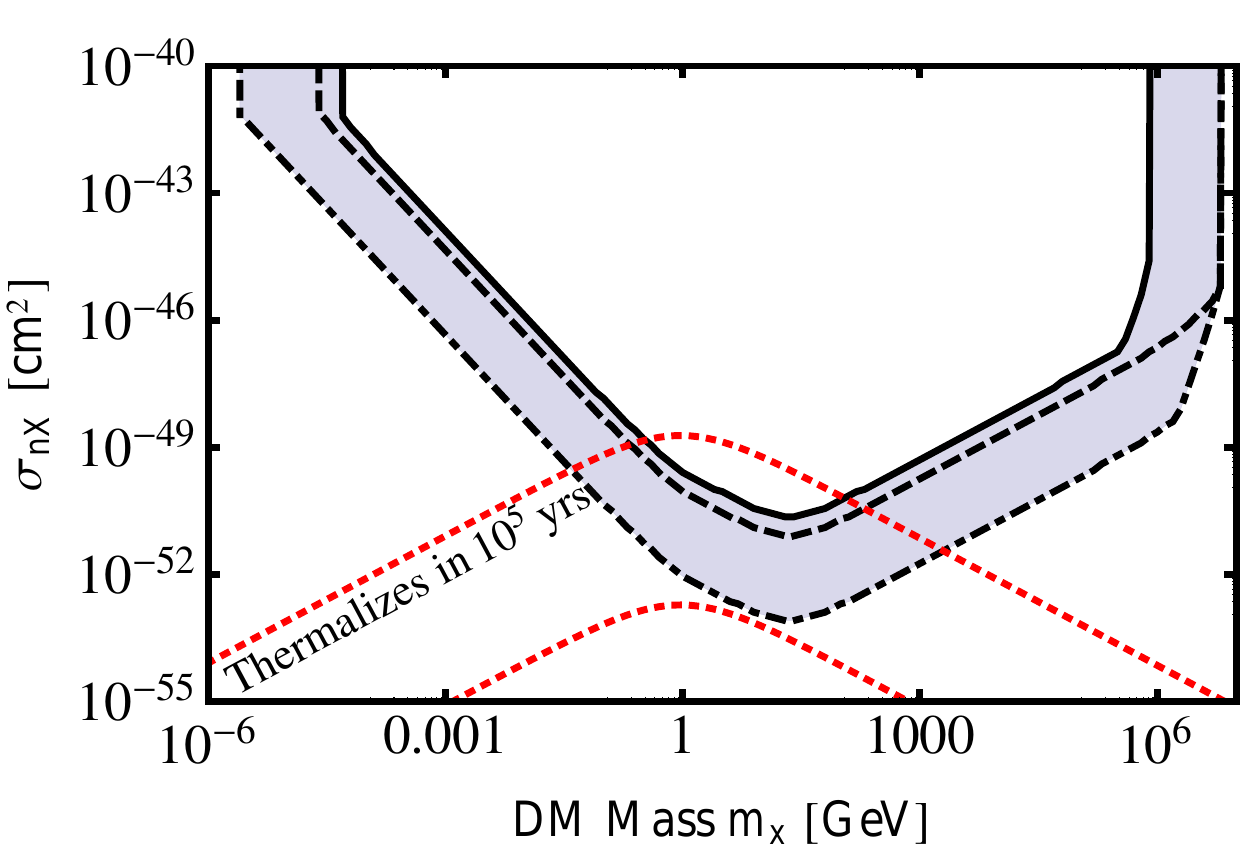}
\\
\includegraphics[scale=.6]{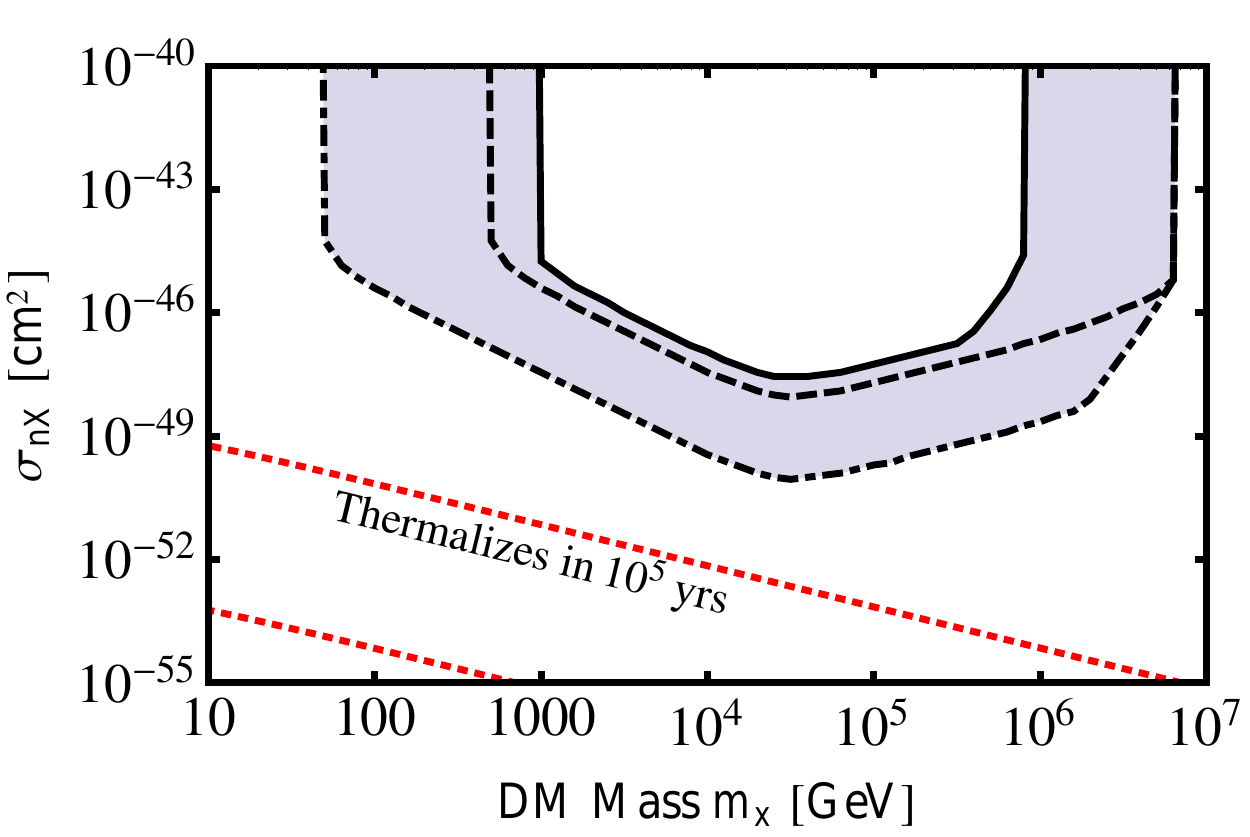}
\end{tabular}
\caption{Bosonic DM nucleon scattering cross-section bounds for pulsar J0437-4715 outside the GC (black solid line), the newly discovered magnetar J1745-2900 (black dashed line), and the lower bound on these parameters assuming DM has collapsed millisecond pulsars $10^7$ years old at the GC (black dot-dashed line) are shown. If bosonic DM has collapsed millisecond pulsars, its parameters will lie in the shaded region between these contours, except for the region below the no thermalization contour (dashed, red), under which DM will not settle into the core of a pulsar. The upper dashed line indicates thermalization within $10^5$ years, the lower, $10^9$ years. The upper panel shows bosonic DM with very small self-interactions ($\lambda =10^{-30}$), while the bottom panel assumes bosonic DM with a small $\lambda |\phi|^4$ term ($\lambda=10^{-15}$).}
\label{fig:boson}
\end{figure}

\begin{figure}
\includegraphics[scale=.6]{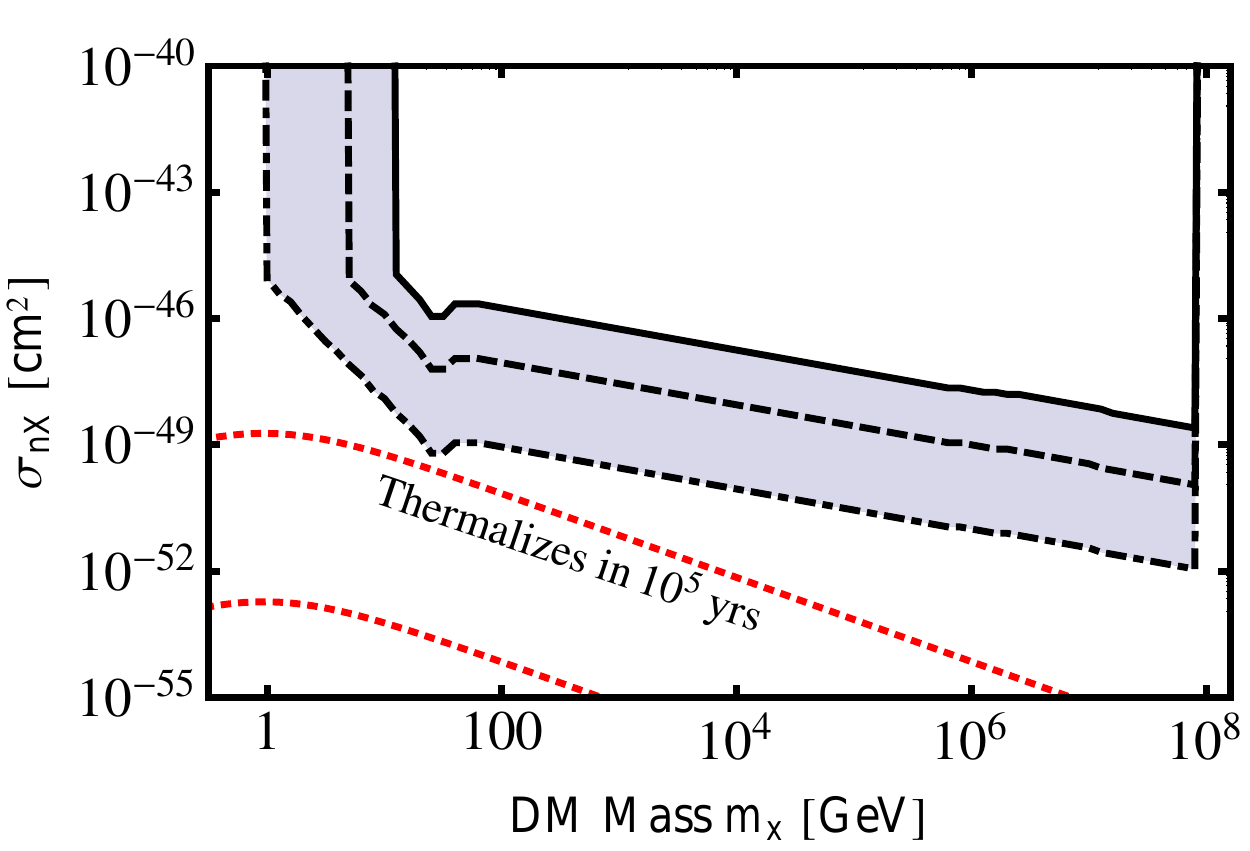}
\caption{Curves for which fermionic DM will destroy a 7 Gyr pulsars near earth, young pulsars 0.1 pc from the GC, and GC millisecond pulsars, are shown with conventions given in Figure \ref{fig:boson}. The parameter space shown is for fermionic DM with a Yukawa coupling ($\alpha=0.1$) to a light ($m_\phi=10~{\rm MeV}$) boson. DM masses $< 100 {~\rm GeV}$ may be excluded by bullet clusters and the ellipticity of spiral galaxies \cite{Tulin:2013teo}.}
\label{fig:fermion}
\end{figure}

In order for the DM captured by a pulsar to collapse into a BH, the DM must clump within a small enough space that its energy is minimized at the Schwarzschild radius. This clumping occurs because the DM rescatters with partons in the pulsar until it has thermalized at a pulsar temperature $\sim 10^5-10^6 ~{\rm K}$. A recent calculation of DM  thermalizing in a non-relativistic Fermi gas via heavy mediators \cite{Bertoni:2013bsa} found a thermalization time of 
\begin{align}
t_{th}\simeq 3.7 ~{\rm kyr} \frac{\frac{m_X}{m_B}}{(1+\frac{m_X}{m_B})^2} \lp \frac{2 \times 10^{-45} ~{\rm cm^2}}{\sigma_{nX}} \rp \lp \frac{10^5 ~{\rm K}}{T_{NS}} \rp^2,
\label{eq:therm}
\end{align}
where $T_{NS}$ is the temperature of the pulsar. This thermalization time substantially impacts pulsar collapse at the GC, as shown in Figures \ref{fig:boson} and \ref{fig:fermion}. The radius of thermalization can be estimated using the virial equation, $r_{th}= \lp \frac{9k_B T_{NS}}{4 \pi G\rho_B m_X} \rp^{1/2}$, where $\rho_B$ is the density of nucleons in the pulsar.

After DM has collected into a tight space in the pulsar, it will collapse to a BH either by forming into a Bose-Einstein condensate (BEC) or degenerate fermions. Turning our attention to the bosonic case \cite{Jamison:2013yya}, a BEC will form out of bosons in excess of $N_{BEC} =  \zeta(3)\lp \frac{k_B T_{NS}}{\sqrt{4 \pi G (\rho_B/3 + P_B)}} \rp^3$, where $ \rho_B $ and $P_B \sim 0.3 \rho_B$ are the core density and pressure of a pulsar \cite{Steiner:2012xt}. 

This DM BEC will collapse to a BH when it passes the Chandrasekhar limit \cite{Bramante:2013hn}. For a bosonic field with a self-coupling term $\lambda |\phi|^4$, the Chandrasekhar limit for bosons is $N_{\rm Chand} = \frac{2 \mpl^2}{\pi m_X^2} \lp1+\frac{{\lambda \mpl^2}}{32 \pi m_X^2} \rp^{1/2}$. The dark BH will consume the pulsar rather than dissipating via Hawking radiation so long as,
\begin{align}
\frac{4 \pi \rho_B (G M_{BH})^2}{v_s^3}  - \frac{1}{15360 \pi (G M_{BH})^2} + C_X m_X>0,
\label{eq:bosonbhgrow}
\end{align}
where $M_{BH} = N_{\rm Chand} m_X$ is the mass of the BH and $v_s/c \sim 0.1$ is the sound speed in a pulsar. The first and second terms of (\ref{eq:bosonbhgrow}) are the Bondi accretion and Hawking radiation, while the third term accounts for DM feeding into the BH from a BEC phase. As detailed in \cite{McDermott:2011jp,Bramante:2013hn} (see also \cite{Kouvaris:2013kra}), for $m_X \gtrsim 100 ~{\rm GeV}$ DM the black hole will grow if subsequent BEC DM falls in more rapidly than the BH evaporates. BEC DM not infalling this rapidly would invalidate the $m_X \gtrsim 100 ~{\rm GeV}$ parameter space in the top panel of Figure \ref{fig:boson}.

Using the recent observation of a young and highly magnetized pulsar near the GC \citep{Mori:2013yda, Kennea:2013dfa, 2013ATel.5043....1E, 2013ATel.5053....1B}, along with the detection of a much older pulsar near Earth with a more diffuse DM background ($\rho_X \sim 0.4 ~ {\rm GeV/cm^3}$, $\bar{v}_X \sim 220 ~ {\rm km/s}$), we can delineate what properties of DM are consistent with dark BHs consuming pulsars at the GC, while pulsars outside the GC remain extant. Pulsar J0437-4715 is a $6.7 ~{\rm Gyr}$ old millisecond pulsar 150 parsecs from Earth \cite{Kargaltsev:2003eb} with a measured surface temperature of $\sim 10^5~{\rm K}$, which implies a core temperature of $\sim 10^6~{\rm K}$ \cite{Manchester:2004bp,McDermott:2011jp}. The newly discovered magnetar J1745-2900, 0.1 pc from the GC, resides in a much denser bath of DM ($\rho_X \sim 7 \times 10^4 ~ {\rm GeV/cm^3}$, $\bar{v}_X \sim 200 ~ {\rm km/s}$). In this work we use typical values for pulsar mass and radius, $1.5 M_\odot \simeq 1.7 \times 10^{57} {~ \rm GeV}$ and $11 ~ {\rm km}$, and assume the magnetar at the GC has a temperature $\sim 10^6~{\rm K}$.

In Figure \ref{fig:boson} we show what DM masses and nucleon scattering cross-sections are consistent with pulsars older than $\sim 10^5-10^7 ~{\rm yr}$ imploding at the GC, while longer-lived pulsars near earth do not collapse. Assuming that absent $\sim {\rm 10^7~ yr}$ old millisecond pulsars at the GC have collapsed into BHs, these simultaneous requirements restrict asymmetric DM to a band of masses and very small scattering cross sections. Of particular interest is the bound on the DM-nucleon cross-section set by the young pulsar, shown as a thick dashed line in Figure \ref{fig:boson}. A future collapse of this young pulsar would indicate that DM mass and couplings lie on this bounding line. For the same reason, if one assumes the absence of expected young pulsars at the GC \cite{Dexter:2013xga} is the result of DM collapse, the DM parameters should lie just below this line. Furthermore, if young pulsars are being destroyed by bosonic DM with very small self-interactions ($\lambda < 10^{-30}$ \citep{Bramante:2013hn}) as in the top panel, the DM responsible cannot have a mass between 10 MeV and a TeV -- this range is ruled out, because thermalization takes longer than $10^5$ years.

We consider DM fermions which collect in a pulsar and collapse by overcoming Fermi degeneracy pressure with very strong attractive self-interactions. We focus on fermionic DM self-coupled by a light scalar mediator, ${V_{Yuk}}=\alpha \phi X \bar{X}$ with mediator mass $m_{\phi} = 10 ~ {\rm MeV}$ and Yukawa coupling $\alpha=0.1$. This Yukawa term is a possible explanation for flattened dwarf galaxy core mass halos \cite{Tulin:2012wi,Tulin:2013teo}. DM fermions with this self-coupling and even very small nucleon scattering cross-sections will collapse in pulsars \cite{Kouvaris:2011gb,Bramante:2013nma}. To determine the number of DM fermions necessary for collapse, we numerically solve for $N_X$ in
\begin{align}
-E_k +{V_{Yuk}^{\rm (Virial)}}+\frac{ G N_X^{2/3}\rho_B m_X}{(\frac{4}{3}\pi)^{-1/3} m_\phi^2} \lp y^2 +\frac{m_Xm_\phi^3}{\rho_B y}  \rp = 0,
\label{eq:fermvirial}
\end{align} 
the virial equation of a DM particle at the edge of the thermalized region, where the virialized kinetic energy $E_k$ is given by $3k_B T_{NS}$ for nondegenerate and $\frac{(3\pi^2)^{2/3}m_\phi^2}{m_Xy^2}$ for degenerate DM. The variable $y\simeq 1.6 m_\phi r/N_X^{1/3}$ is the exponent of the Yukawa potential for nearest neighbor fermions. Note that at collapse, the self-gravity of the fermions is negligible. In this treatment we only consider DM parameters for which the Yukawa potential is strongly-screened ($y>1$) at the onset of collapse, so that in (\ref{eq:fermvirial}) ${V_{Yuk}^{\rm (Virial)}} \simeq 8\alpha\lp \frac{m_\phi e^{-y}}{y}+m_\phi e^{-y} \rp$. The last term in (\ref{eq:fermvirial}) is the baryonic and DM gravitational potential.

If $\alpha > 4.7 (m_\phi/m_X)^2$ collapse will continue when the DM becomes relativistic. More details can be found in \cite{Bramante:2013nma}. Unlike the bosonic case in (\ref{eq:bosonbhgrow}), the DM fermions are not confined to a BEC and so will not efficiently feed into the BH after it is formed. The pulsar will be swallowed if the number of collapsing fermions exceeds $N_X \sim \lp \frac{3.4 \times 10^{36} ~ {\rm GeV}}{m_X} \rp$, which is the number required for the baryonic accretion of the BH to outpace the Hawking radiation rate. In Figure \ref{fig:fermion} we plot a detection band for self-interacting fermionic DM which collapses pulsars in the GC.

\begin{figure}
\includegraphics[scale=0.6]{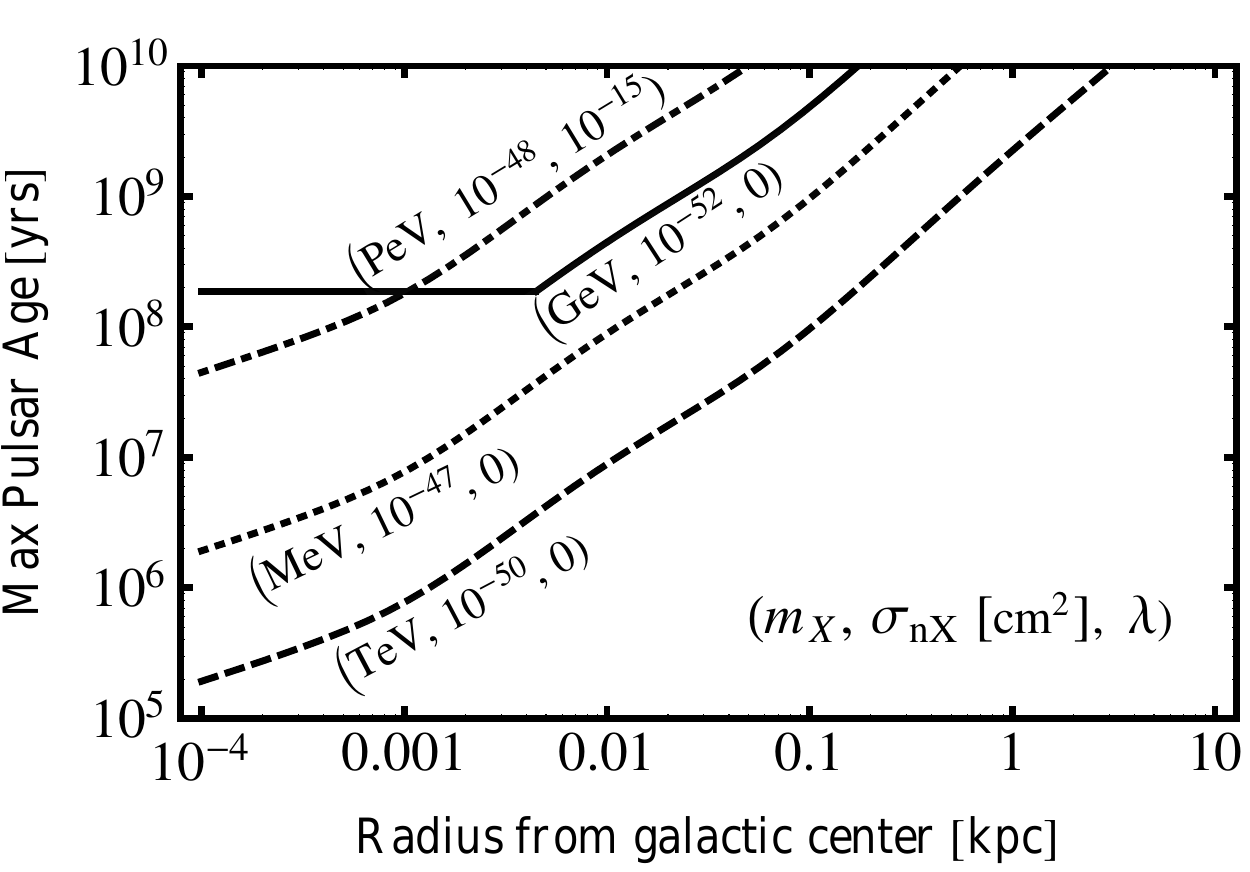}
\caption{This figure displays maximum pulsar age as a function of distance from the GC for a number of asymmetric bosonic DM models. The model for each curve is labeled in the format (Mass, $\sigma_{nX}~ $(${\rm cm^2}$), $\lambda$). The flat pulsar age distribution predicted for GeV DM (solid black line) is a signal that would appear for asymmetric bosonic DM in the 10 MeV - TeV range as a result of longer thermalization times in pulsars.}
\label{fig:radial}
\end{figure}

A prediction of GC pulsar collapsing DM is that there is a maximum age that a pulsar can reach before a DM BH devours the star. This maximum age should depend sensitively on the local DM density (\ref{eq:scattering}), and thus on the distance of the pulsar from the center of the Milky Way. The shape and normalization of this maximum age curve could be used to help determine DM mass, couplings, and self-interactions. Efforts to precisely determine the mass distribution and velocity profile of the inner parsec are ongoing, however there is ample evidence that baryonic matter at 0.1 parsecs achieves velocities $\sim 200 {\rm ~ km/s }$ \cite{Ghez:2003qj,Lu:2008iz,Kaplinghat:2013xca}. In this study we assume that the DM density follows an NFW profile
$
\rho (r) = \rho_0 \frac{\lp r/r_s\rp^{-\gamma}}{\lp 1+r/r_s \rp^{3-\gamma}},
$
with $\gamma=1.0$, $r_s=20 {~\rm kpc}$, and $\rho_0 = {0.4 ~{\rm GeV/cm^3}}$ \cite{Navarro:1996gj,Iocco:2011jz}. A less-cusped profile would decrease GC DM density, leading to longer pulsar collapse times. To fit galactic velocity profiles both outside $r\gtrsim 0.5 {~ \rm pc}$ \cite{2013PASJ...65..118S} along with the observed velocities of stars in the central parsec \cite{Kaplinghat:2013xca}, we assume a mass distribution given by 
\begin{align}
M(r) =M_0+ 4 \pi \int_0^r\left[ \rho(r_1) +\rho_{H1}(r_1) +\rho_{H2}(r_1) \right] r_1^2~ dr_1,
\end{align} 
where the first term is $M_0 = 10^{63} {\rm ~ GeV}$, and the last two density terms are Hernquist potentials, $\rho_{Hi} (r)= \rho_{H0i} (r^{-1})/(r_{0i}+r)^3 $, with $\rho_{H01} = 2\times10^3 {~\rm GeV/cm^3}$, $r_{01}=2.7 {\rm~kpc}$, $\rho_{H02} = 4 {~ \rm MeV/cm^3}$ and $r_{02}=0.01 {~\rm kpc}$. We approximate the velocity dispersion as the orbital velocity, $\bar{v}(r)\sim v_c(r) = \sqrt{G M(r)/r}$.

In Figure \ref{fig:radial} we plot curves of maximum pulsar age as a function of distance from the GC. Such a dependence would be a signal of asymmetric DM. The coupling of DM to nucleons determines the normalization of this curve. Curves with a higher normalization destroy only millisecond pulsars, while lower curves destroy young pulsars at the GC (this implies weaker and stronger nucleon scattering, respectively). It is intriguing that the detection of a flat pulsar age distribution in the central parsec would imply bosonic DM with a mass between 10 MeV and a TeV and a very weak quartic coupling ($\lambda =0$). DM with a mass and quartic coupling in this range, because of the tiny nucleon scattering it is required to have vis-a-vis solar position pulsar bounds, cannot scatter efficiently with neutrons and will not have enough time to clump into a BH progenitor.  

Our results show that asymmetric DM can collapse old MSPs at the GC, remaining consistent with local MSP observations. The ``missing pulsar problem" also extends to young pulsars -- we have shown these may be converted into BHs $\sim 10^5$ years after forming. Interestingly, the one observed pulsar in the GC region is a magnetar, and thus probably one of the youngest pulsars in our galaxy. While this may indicate that young pulsars are destroyed by DM near the GC, it may also indicate more routine astrophysical phenomena, like a slow star formation rate between 25 -- 60 Myr ago, or a top-heavy stellar initial mass function that preferences the formation of BHs over pulsars~\citep{Nayakshin:2005re}. Additionally, it is worth noting that while several low mass X-ray binaries are observed near the GC it is not known whether the compact object in these systems is currently a pulsar or BH~\citep{Muno:2004xi}. Direct collapse of a pulsar is unlikely to disrupt a stable low-mass X-ray binary.

Some prior work on DM NS bounds speculated that a nearby Gyr old pulsar in globular cluster M4 may reside in a DM background density of $\sim 10^3 {\rm ~ GeV/cm^3}$ \citep{McCullough:2010ai,McDermott:2011jp,Bramante:2013hn}. Such a pulsar is nominally at odds with some of the DM parameter space presented here. However, as noted by the same papers, M4 is unlikely to have a high DM density \cite{Gilmore:2007fy,Wolf:2009tu}.

On the other hand, if old pulsars are found to collapse via DM accumulation in the GC, with old pulsars in dense DM halos elsewhere, this would be an indication of multicomponent DM. Indeed, multicomponent DM frameworks with varied spatial and temporal abundances \cite{Chen:2006ni,Dienes:2011ja,Dienes:2011sa,Fan:2013yva,Fan:2013tia} could explain how DM at the GC destroys pulsars while other DM does not. Particularly, a disk of asymmetric DM \cite{Fan:2013bea} with sizable nucleon scattering would explain old pulsars outside the disk plane surviving. A pulsar signal of multicomponent DM could be correlated with other signals of multicomponent DM, for example a DM decay mass spectrum in AMS-02 cosmic-ray positron fraction measurements \cite{Dienes:2013lxa}. In fact, pulsar collapsing DM can have significant decays without upsetting the collapse mechanism \cite{Bramante:2013hn}. This is a particular strength of searching for pulsar collapses due to DM accumulation -- they provide a handle on the spatial distribution of asymmetric DM, in contrast with conventional indirect methods for studying annihilating DM. Furthermore, pulsar age searches are sensitive to nucleon scattering cross sections well below those attainable with conventional direct detection methods. We leave a complete study of multicomponent pulsar collapsing DM, and other applications of pulsar-crushing DM in the central parsec to future work.

{\bf Acknowledgements.}
We thank Adam Martin for useful comments on the manuscript, and Eric Carlson, Keith Dienes, JiJi Fan, Patrick Fox, Dan Hooper, Gordan Krnjaic, James Unwin, and Brooks Thomas for discussions. In addition, we wish to thank Fermilab and the organizers of the New Perspectives on Dark Matter Workshop for their hospitality while this work was completed.

\bibliographystyle{h-physrev.bst}

\bibliography{dmpulsars}

\begin{thebibliography}{10}

\bibitem{Hooper:2010mq}
D.~Hooper and L.~Goodenough,
\newblock Phys.Lett. {\bf B697}, 412 (2011), 1010.2752.

\bibitem{Abazajian:2012pn}
K.~N. Abazajian and M.~Kaplinghat,
\newblock Phys.Rev. {\bf D86}, 083511 (2012), 1207.6047.

\bibitem{Gordon:2013vta}
C.~Gordon and O.~Macias,
\newblock Phys.Rev. {\bf D88}, 083521 (2013), 1306.5725.

\bibitem{Daylan:2014rsa}
T.~Daylan {\em et~al.},
\newblock (2014), 1402.6703.

\bibitem{Bulbul:2014sua}
E.~Bulbul {\em et~al.},
\newblock (2014), 1402.2301.

\bibitem{Boyarsky:2014jta}
A.~Boyarsky, O.~Ruchayskiy, D.~Iakubovskyi, and J.~Franse,
\newblock (2014), 1402.4119.

\bibitem{Rocha:2012jg}
M.~Rocha {\em et~al.},
\newblock Mon.Not.Roy.Astron.Soc. {\bf 430}, 81 (2013), 1208.3025.

\bibitem{Peter:2012jh}
A.~H. Peter, M.~Rocha, J.~S. Bullock, and M.~Kaplinghat,
\newblock (2012), 1208.3026.

\bibitem{Vogelsberger:2012ku}
M.~Vogelsberger, J.~Zavala, and A.~Loeb,
\newblock Mon.Not.Roy.Astron.Soc. {\bf 423}, 3740 (2012), 1201.5892.

\bibitem{Zavala:2012us}
J.~Zavala, M.~Vogelsberger, and M.~G. Walker,
\newblock Monthly Notices of the Royal Astronomical Society: Letters {\bf 431},
  L20 (2013), 1211.6426.

\bibitem{Mori:2013yda}
K.~Mori {\em et~al.},
\newblock Astrophys.J. {\bf 770}, L23 (2013), 1305.1945.

\bibitem{Kennea:2013dfa}
J.~Kennea {\em et~al.},
\newblock Astrophys.J. {\bf 770}, L24 (2013), 1305.2128.

\bibitem{2013ATel.5043....1E}
R.~{Eatough} {\em et~al.},
\newblock The Astronomer's Telegram {\bf 5043}, 1 (2013).

\bibitem{2013ATel.5053....1B}
M.~{Buttu} {\em et~al.},
\newblock The Astronomer's Telegram {\bf 5053}, 1 (2013).

\bibitem{Spitler:2013uva}
L.~Spitler {\em et~al.},
\newblock Astrophys.J. {\bf 780}, L3 (2014), 1309.4673.

\bibitem{Bower:2013tva}
G.~C. Bower {\em et~al.},
\newblock Astrophys.J. {\bf 780}, L2 (2014), 1309.4672.

\bibitem{Pfahl:2003tf}
E.~Pfahl and A.~Loeb,
\newblock Astrophys.J. {\bf 615}, 253 (2004), astro-ph/0309744.

\bibitem{Macquart:2010vf}
J.-P. Macquart, N.~Kanekar, D.~Frail, and S.~Ransom,
\newblock Astrophys.J. {\bf 715}, 939 (2010), 1004.1643.

\bibitem{Dexter:2013xga}
J.~Dexter and R.~M. O'Leary,
\newblock Astrophys.J. {\bf 783}, L7 (2014), 1310.7022.

\bibitem{Chennamangalam:2013zja}
J.~Chennamangalam and D.~Lorimer,
\newblock (2013), 1311.4846.

\bibitem{Goldman:1989nd}
I.~Goldman and S.~Nussinov,
\newblock Phys.Rev. {\bf D40}, 3221 (1989).

\bibitem{Kouvaris:2007ay}
C.~Kouvaris,
\newblock Phys.Rev. {\bf D77}, 023006 (2008), 0708.2362.

\bibitem{McCullough:2010ai}
M.~McCullough and M.~Fairbairn,
\newblock Phys.Rev. {\bf D81}, 083520 (2010), 1001.2737.

\bibitem{deLavallaz:2010wp}
A.~de~Lavallaz and M.~Fairbairn,
\newblock Phys.Rev. {\bf D81}, 123521 (2010), 1004.0629.

\bibitem{Kouvaris:2010vv}
C.~Kouvaris and P.~Tinyakov,
\newblock Phys.Rev. {\bf D82}, 063531 (2010), 1004.0586.

\bibitem{Kouvaris:2010jy}
C.~Kouvaris and P.~Tinyakov,
\newblock Phys.Rev. {\bf D83}, 083512 (2011), 1012.2039.

\bibitem{McDermott:2011jp}
S.~D. McDermott, H.-B. Yu, and K.~M. Zurek,
\newblock Phys.Rev. {\bf D85}, 023519 (2012), 1103.5472.

\bibitem{Kouvaris:2011fi}
C.~Kouvaris and P.~Tinyakov,
\newblock Phys.Rev.Lett. {\bf 107}, 091301 (2011), 1104.0382.

\bibitem{Kouvaris:2011gb}
C.~Kouvaris,
\newblock Phys.Rev.Lett. {\bf 108}, 191301 (2012), 1111.4364.

\bibitem{Casanellas:2012jp}
J.~Casanellas and I.~Lopes,
\newblock Astrophys.J. {\bf 765}, L21 (2013), 1212.2985.

\bibitem{Bramante:2013hn}
J.~Bramante, K.~Fukushima, and J.~Kumar,
\newblock Phys.Rev. {\bf D87}, 055012 (2013), 1301.0036.

\bibitem{Bell:2013xk}
N.~F. Bell, A.~Melatos, and K.~Petraki,
\newblock Phys.Rev. {\bf D87}, 123507 (2013), 1301.6811.

\bibitem{Goldman:2013qla}
I.~Goldman, R.~Mohapatra, S.~Nussinov, D.~Rosenbaum, and V.~Teplitz,
\newblock Phys.Lett. {\bf B725}, 200 (2013), 1305.6908.

\bibitem{Jamison:2013yya}
A.~O. Jamison,
\newblock Phys.Rev. {\bf D88}, 035004 (2013), 1304.3773.

\bibitem{Bertoni:2013bsa}
B.~Bertoni, A.~E. Nelson, and S.~Reddy,
\newblock Phys.Rev. {\bf D88}, 123505 (2013), 1309.1721.

\bibitem{Bramante:2013nma}
J.~Bramante, K.~Fukushima, J.~Kumar, and E.~Stopnitzky,
\newblock Phys.Rev. {\bf D89}, 015010 (2014), 1310.3509.

\bibitem{Kouvaris:2013kra}
C.~Kouvaris and P.~Tinyakov,
\newblock (2013), 1312.3764.

\bibitem{Petraki:2013wwa}
K.~Petraki and R.~R. Volkas,
\newblock Int.J.Mod.Phys. {\bf A28}, 1330028 (2013), 1305.4939.

\bibitem{Zurek:2013wia}
K.~M. Zurek,
\newblock Phys.Rept. {\bf 537}, 91 (2014), 1308.0338.

\bibitem{Gould:1987ir}
A.~Gould,
\newblock Astrophys.J. {\bf 321}, 571 (1987).

\bibitem{Steiner:2012xt}
A.~W. Steiner, J.~M. Lattimer, and E.~F. Brown,
\newblock Astrophys.J. {\bf 765}, L5 (2013), 1205.6871.

\bibitem{Kargaltsev:2003eb}
O.~Kargaltsev, G.~Pavlov, and R.~Romani,
\newblock Astrophys.J. {\bf 602}, 327 (2004), astro-ph/0310854.

\bibitem{Manchester:2004bp}
R.~N. Manchester, G.~B. Hobbs, A.~Teoh, and M.~Hobbs,
\newblock Astron.J. {\bf 129}, 1993 (2005), astro-ph/0412641.

\bibitem{Tulin:2013teo}
S.~Tulin, H.-B. Yu, and K.~M. Zurek,
\newblock (2013), 1302.3898.

\bibitem{Tulin:2012wi}
S.~Tulin, H.-B. Yu, and K.~M. Zurek,
\newblock (2012), 1210.0900.

\bibitem{Ghez:2003qj}
A.~Ghez {\em et~al.},
\newblock Astrophys.J. {\bf 620}, 744 (2005), astro-ph/0306130.

\bibitem{Lu:2008iz}
J.~Lu {\em et~al.},
\newblock Astrophys.J. {\bf 690}, 1463 (2009), 0808.3818.

\bibitem{Kaplinghat:2013xca}
M.~Kaplinghat, R.~E. Keeley, T.~Linden, and H.-B. Yu,
\newblock (2013), 1311.6524.

\bibitem{Navarro:1996gj}
J.~F. Navarro, C.~S. Frenk, and S.~D. White,
\newblock Astrophys.J. {\bf 490}, 493 (1997), astro-ph/9611107.

\bibitem{Iocco:2011jz}
F.~Iocco, M.~Pato, G.~Bertone, and P.~Jetzer,
\newblock JCAP {\bf 1111}, 029 (2011), 1107.5810.

\bibitem{2013PASJ...65..118S}
Y.~{Sofue},
\newblock  {\bf 65}, 118 (2013), 1307.8241.

\bibitem{Nayakshin:2005re}
S.~Nayakshin and R.~Sunyaev,
\newblock Mon.Not.Roy.Astron.Soc.Lett. {\bf 364}, L23 (2005), astro-ph/0507687.

\bibitem{Muno:2004xi}
M.~P. Muno {\em et~al.},
\newblock Astrophys.J. {\bf 622}, L113 (2005), astro-ph/0412492.

\bibitem{Gilmore:2007fy}
G.~Gilmore {\em et~al.},
\newblock Astrophys.J. {\bf 663}, 948 (2007), astro-ph/0703308.

\bibitem{Wolf:2009tu}
J.~Wolf {\em et~al.},
\newblock Mon.Not.Roy.Astron.Soc. {\bf 406}, 1220 (2010), 0908.2995.

\bibitem{Chen:2006ni}
X.~Chen and S.-H.~H. Tye,
\newblock JCAP {\bf 0606}, 011 (2006), hep-th/0602136.

\bibitem{Dienes:2011ja}
K.~R. Dienes and B.~Thomas,
\newblock Phys.Rev. {\bf D85}, 083523 (2012), 1106.4546.

\bibitem{Dienes:2011sa}
K.~R. Dienes and B.~Thomas,
\newblock Phys.Rev. {\bf D85}, 083524 (2012), 1107.0721.

\bibitem{Fan:2013yva}
J.~Fan, A.~Katz, L.~Randall, and M.~Reece,
\newblock Phys.Dark Univ. {\bf 2}, 139 (2013), 1303.1521.

\bibitem{Fan:2013tia}
J.~Fan, A.~Katz, L.~Randall, and M.~Reece,
\newblock Phys.Rev.Lett. {\bf 110}, 211302 (2013), 1303.3271.

\bibitem{Fan:2013bea}
J.~Fan, A.~Katz, and J.~Shelton,
\newblock (2013), 1312.1336.

\bibitem{Dienes:2013lxa}
K.~R. Dienes, J.~Kumar, and B.~Thomas,
\newblock Phys.Rev. {\bf D88}, 103509 (2013), 1306.2959.

\end{thebibliography}

\end{document}